\pdfoutput=1
\documentclass[12pt,a4paper]{article}

\usepackage{ifthen} 
\usepackage{placeins}
\usepackage{floatrow}
\usepackage{booktabs}

\newboolean{pdflatex}
\setboolean{pdflatex}{true} 

\newboolean{articletitles}
\setboolean{articletitles}{true} 

\newboolean{uprightparticles}
\setboolean{uprightparticles}{false} 

\def\paperauthors{LHCb collaboration} 
\def\paperasciititle{Enhanced production of Lambda_b baryons in high-multiplicity pp collisions at sqrt(s) = 13 TeV} 
\def\papertitle{Enhanced production of $\Lb$ baryons in high-multiplicity $pp$ collisions at $\sqs = 13\tev$}
\def\paperkeywords{{High Energy Physics}, {LHCb}} 
\def\papercopyright{\the\year\ CERN for the benefit of the LHCb collaboration} 
\def\paperlicence{CC BY 4.0 licence}
\def\paperlicenceurl{https://creativecommons.org/licenses/by/4.0/}

\usepackage[top=1in, bottom=1.25in, left=1in, right=1in]{geometry}

\usepackage{microtype}
\usepackage{lineno}  
\usepackage{xspace} 
\usepackage{caption} 

\usepackage{graphicx}  
\usepackage{color}
\usepackage{colortbl}
\graphicspath{{./figs/}} 

\usepackage{amsmath} 
\usepackage{amssymb}
\usepackage{amsfonts}
\usepackage{upgreek} 

\newcommand*\patchAmsMathEnvironmentForLineno[1]{%
\expandafter\let\csname old#1\expandafter\endcsname\csname #1\endcsname
\expandafter\let\csname oldend#1\expandafter\endcsname\csname
end#1\endcsname
 \renewenvironment{#1}%
   {\linenomath\csname old#1\endcsname}%
   {\csname oldend#1\endcsname\endlinenomath}%
}
\newcommand*\patchBothAmsMathEnvironmentsForLineno[1]{%
  \patchAmsMathEnvironmentForLineno{#1}%
  \patchAmsMathEnvironmentForLineno{#1*}%
}
\AtBeginDocument{%
\patchBothAmsMathEnvironmentsForLineno{equation}%
\patchBothAmsMathEnvironmentsForLineno{align}%
\patchBothAmsMathEnvironmentsForLineno{flalign}%
\patchBothAmsMathEnvironmentsForLineno{alignat}%
\patchBothAmsMathEnvironmentsForLineno{gather}%
\patchBothAmsMathEnvironmentsForLineno{multline}%
\patchBothAmsMathEnvironmentsForLineno{eqnarray}%
}

\usepackage{hyperxmp}

\usepackage[pdftex,
            pdfauthor={\paperauthors},
            pdftitle={\paperasciititle},
            pdfkeywords={\paperkeywords},
            pdfcopyright={Copyright (C) \papercopyright},
            pdflicenseurl={\paperlicenceurl}]{hyperref}

\usepackage[colorinlistoftodos,textsize=scriptsize]{todonotes}

\usepackage[bottom,flushmargin,hang,multiple]{footmisc}

\usepackage[all]{hypcap} 

\usepackage{xspace} 
\usepackage{upgreek}

\def\lhcb   {\mbox{LHCb}\xspace}

\def\MagUp {\mbox{\em Mag\kern -0.05em Up}\xspace}

\ifthenelse{\boolean{uprightparticles}}%
{

 \def\Pmu         {\ensuremath{\upmu}\xspace}

 \def\Ppsi        {\ensuremath{\uppsi}\xspace}

 \def\PDelta      {\ensuremath{\Delta}\xspace}                 
 \def\PXi         {\ensuremath{\Xi}\xspace}                 
 \def\PLambda     {\ensuremath{\Lambda}\xspace}                 
 \def\PSigma      {\ensuremath{\Sigma}\xspace}                 
 \def\POmega      {\ensuremath{\Omega}\xspace}                 
 \def\PUpsilon    {\ensuremath{\Upsilon}\xspace}
 \let\oldPi\Pi
 \def\PPi         {\ensuremath{\oldPi}\xspace}

 \def\PB      {\ensuremath{\mathrm{B}}\xspace}                 
                  
 \def\PD      {\ensuremath{\mathrm{D}}\xspace}

 \def\PJ      {\ensuremath{\mathrm{J}}\xspace}                 
 \def\PK      {\ensuremath{\mathrm{K}}\xspace}

 \def\PZ      {\ensuremath{\mathrm{Z}}\xspace}                 
                  
 \def\Pb      {\ensuremath{\mathrm{b}}\xspace}                 
 \def\Pc      {\ensuremath{\mathrm{c}}\xspace}

 \def\Pi      {\ensuremath{\mathrm{i}}\xspace}

 \def\Ps      {\ensuremath{\mathrm{s}}\xspace}

 \def\thebaroffset{0.0em}
}
{

 \def\Pmu         {\ensuremath{\mu}\xspace}

 \def\Ppsi        {\ensuremath{\psi}\xspace}                 
                  
 \mathchardef\PDelta="7101
 \mathchardef\PXi="7104
 \mathchardef\PLambda="7103
 \mathchardef\PSigma="7106
 \mathchardef\POmega="710A
 \mathchardef\PUpsilon="7107
 \mathchardef\PPi="7105
                  
 \def\PB      {\ensuremath{B}\xspace}                 
                  
 \def\PD      {\ensuremath{D}\xspace}

 \def\PJ      {\ensuremath{J}\xspace}                 
 \def\PK      {\ensuremath{K}\xspace}

 \def\PZ      {\ensuremath{Z}\xspace}                 
                  
 \def\Pb      {\ensuremath{b}\xspace}                 
 \def\Pc      {\ensuremath{c}\xspace}

 \def\Pi      {\ensuremath{i}\xspace}

 \def\Ps      {\ensuremath{s}\xspace}

 \def\thebaroffset{0.18em}
}
\newcommand{\offsetoverline}[2][\thebaroffset]{\kern #1\overline{\kern -#1 #2}}%

\makeatletter
\ifcase \@ptsize \relax
  \newcommand{\miniscule}{\@setfontsize\miniscule{4}{5}}
\or
  \newcommand{\miniscule}{\@setfontsize\miniscule{5}{6}}
\or
  \newcommand{\miniscule}{\@setfontsize\miniscule{5}{6}}
\fi
\makeatother

\DeclareRobustCommand{\optbar}[1]{\shortstack{{\miniscule (\rule[.5ex]{1.25em}{.18mm})}
  \\ [-.7ex] $#1$}}

\def\mup        {{\ensuremath{\Pmu^+}}\xspace}

\def\Z      {{\ensuremath{\PZ}}\xspace}

\def\squark    {{\ensuremath{\Ps}}\xspace}

\def\cquark    {{\ensuremath{\Pc}}\xspace}

\def\bquark    {{\ensuremath{\Pb}}\xspace}

\def\KorKbar {\kern \thebaroffset\optbar{\kern -\thebaroffset \PK}{}\xspace}

\def\D       {{\ensuremath{\PD}}\xspace}

\def\DorDbar {\kern \thebaroffset\optbar{\kern -\thebaroffset \PD}\xspace}

\def\Dp      {{\ensuremath{\D^+}}\xspace}
\def\Dm      {{\ensuremath{\D^-}}\xspace}

\def\DpDm    {\ensuremath{\Dp {\kern -0.16em \Dm}}\xspace}

\def\B       {{\ensuremath{\PB}}\xspace}

\def\BorBbar {\kern \thebaroffset\optbar{\kern -\thebaroffset \PB}\xspace}

\def\Bd      {{\ensuremath{\B^0}}\xspace}

\def\BdorBdbar {\kern \thebaroffset\optbar{\kern -\thebaroffset \Bd}\xspace}
\def\Bu      {{\ensuremath{\B^+}}\xspace}

\def\Bs      {{\ensuremath{\B^0_\squark}}\xspace}

\def\BsorBsbar {\kern \thebaroffset\optbar{\kern -\thebaroffset \Bs}\xspace}

\def\jpsi     {{\ensuremath{{\PJ\mskip -3mu/\mskip -2mu\Ppsi}}}\xspace}

\def\Y#1S{\ensuremath{\PUpsilon{(#1S)}}\xspace}

\def\Lz          {{\ensuremath{\PLambda}}\xspace}
\def\Lbar        {{\ensuremath{\offsetoverline{\PLambda}}}\xspace}
\def\LorLbar     {\kern \thebaroffset\optbar{\kern -\thebaroffset \PLambda}\xspace}

\def\Lc          {{\ensuremath{\Lz^+_\cquark}}\xspace}

\def\Lb           {{\ensuremath{\Lz^0_\bquark}}\xspace}
\def\Lbbar        {{\ensuremath{\Lbar{}^0_\bquark}}\xspace}

\def\BF         {{\ensuremath{\mathcal{B}}}\xspace}
\def\BR         {\BF}

\def\AT#1     {\ensuremath{A_{\mathrm{T}}^{#1}}\xspace}           

\def\C#1      {\ensuremath{\mathcal{C}_{#1}}\xspace}                       
\def\Cp#1     {\ensuremath{\mathcal{C}_{#1}^{'}}\xspace}                    
\def\Ceff#1   {\ensuremath{\mathcal{C}_{#1}^{\mathrm{(eff)}}}\xspace}        
\def\Cpeff#1  {\ensuremath{\mathcal{C}_{#1}^{'\mathrm{(eff)}}}\xspace}       
\def\Ope#1    {\ensuremath{\mathcal{O}_{#1}}\xspace}                       
\def\Opep#1   {\ensuremath{\mathcal{O}_{#1}^{'}}\xspace}                    

\newcommand{\aunit}[1]{\ensuremath{\text{\,#1}}}       

\newcommand{\tev}{\aunit{Te\kern -0.1em V}\xspace}
\newcommand{\gev}{\aunit{Ge\kern -0.1em V}\xspace}
\newcommand{\mev}{\aunit{Me\kern -0.1em V}\xspace}
\newcommand{\kev}{\aunit{ke\kern -0.1em V}\xspace}
\newcommand{\ev}{\aunit{e\kern -0.1em V}\xspace}
 
\newcommand{\mevc}{\ensuremath{\aunit{Me\kern -0.1em V\!/}c}\xspace}
\newcommand{\gevc}{\ensuremath{\aunit{Ge\kern -0.1em V\!/}c}\xspace}
\newcommand{\mevcc}{\ensuremath{\aunit{Me\kern -0.1em V\!/}c^2}\xspace}
\newcommand{\gevcc}{\ensuremath{\aunit{Ge\kern -0.1em V\!/}c^2}\xspace}

\def\gsim{{~\raise.15em\hbox{$>$}\kern-.85em
          \lower.35em\hbox{$\sim$}~}\xspace}
\def\lsim{{~\raise.15em\hbox{$<$}\kern-.85em
          \lower.35em\hbox{$\sim$}~}\xspace}

\def\sPlot{\mbox{\em sPlot}\xspace}

\def\sqs   {\ensuremath{\protect\sqrt{s}}\xspace}

\def\pt         {\ensuremath{p_{\mathrm{T}}}\xspace}

\def\evtgen     {\mbox{\textsc{EvtGen}}\xspace}

\def\geant      {\mbox{\textsc{Geant4}}\xspace}

\def\photos     {\mbox{\textsc{Photos}}\xspace}

\def\pythia     {\mbox{\textsc{Pythia}}\xspace}

\def\tell1  {TELL1\xspace}
\def\ukl1   {UKL1\xspace}

\newcommand{\lhcborcid}[1]{\href{https://orcid.org/#1}{\hspace*{0.1em}\raisebox{-0.45ex}{\includegraphics[width=1em]{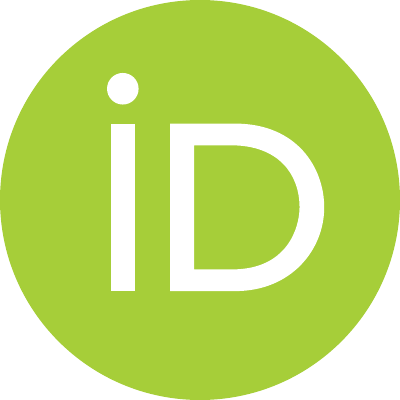}}}}

\usepackage{cite} 
\usepackage{mciteplus}

\def\Nvelo {\ensuremath{\mathrm{N}_{\mathrm{tracks}}^{\mathrm{VELO}}}\xspace}
\def\Nback {\ensuremath{\mathrm{N}_{\mathrm{tracks}}^{\mathrm{back}}}\xspace}

\usepackage{longtable} 

\begin{document}

\renewcommand{\thefootnote}{\fnsymbol{footnote}}
\setcounter{footnote}{1}


\begin{titlepage}
\pagenumbering{roman}

\vspace*{-1.5cm}
\centerline{\large EUROPEAN ORGANIZATION FOR NUCLEAR RESEARCH (CERN)}
\vspace*{1.5cm}
\noindent
\begin{tabular*}{\linewidth}{lc@{\extracolsep{\fill}}r@{\extracolsep{0pt}}}
\ifthenelse{\boolean{pdflatex}}
{\vspace*{-1.5cm}\mbox{\!\!\!\includegraphics[width=.14\textwidth]{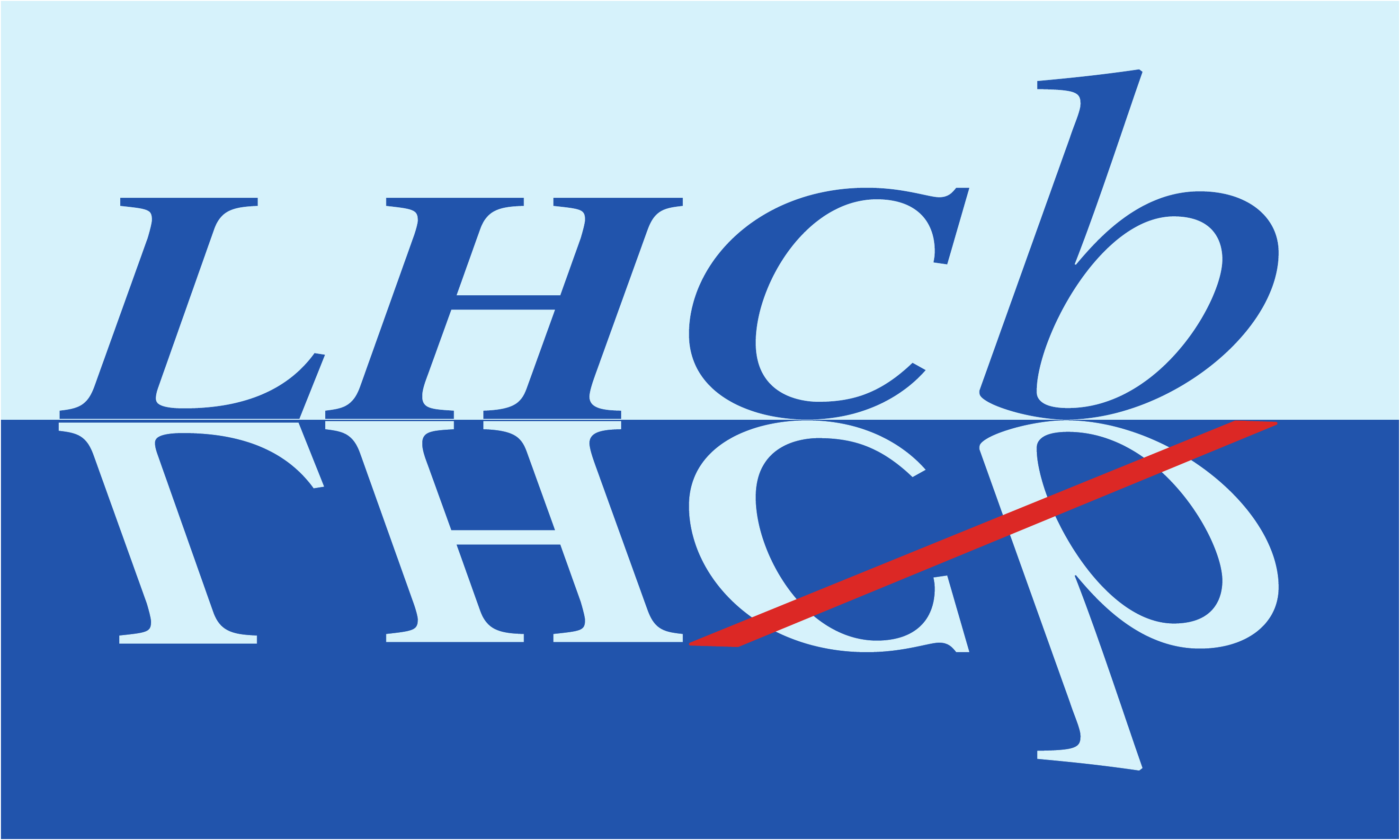}} & &}%
{\vspace*{-1.2cm}\mbox{\!\!\!\includegraphics[width=.12\textwidth]{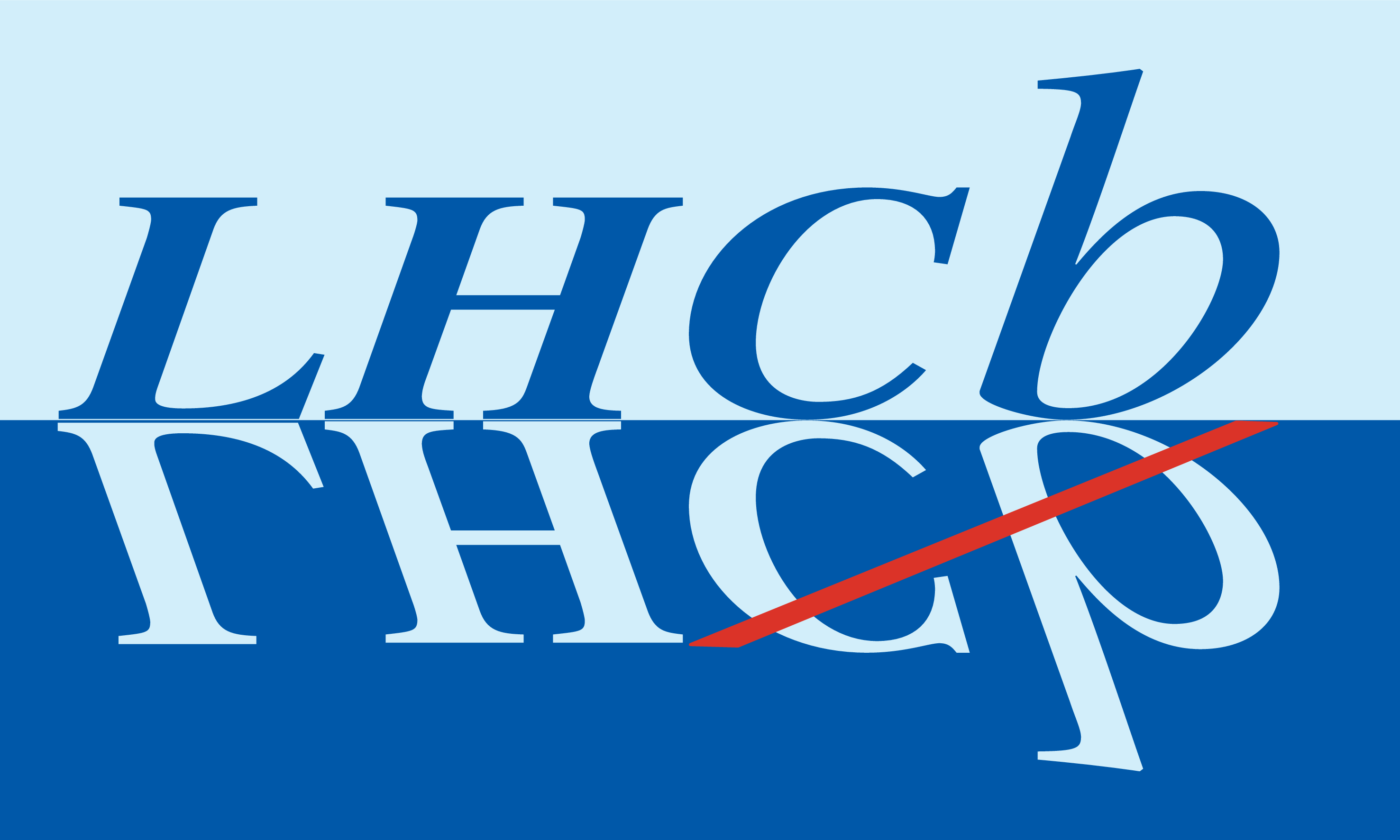}} & &}%
\\
 & & CERN-EP-2023-208 \\  
 & & LHCb-PAPER-2023-027 \\  
 & & \today \\ 
 & & \\
\end{tabular*}

\vspace*{4.0cm}

{\normalfont\bfseries\boldmath\huge
\begin{center}
  \papertitle 
\end{center}
}

\vspace*{2.0cm}
\begin{center}
\paperauthors
\end{center}

\vspace{\fill}

\begin{abstract}
  \noindent
The production rate of $\Lb$ baryons relative to $\Bd$ mesons in $pp$ collisions at a center-of-mass energy $\sqrt{s} = 13 \tev$ is measured by the LHCb experiment.  The ratio of $\Lb$ to $\Bd$ production cross-sections shows a significant dependence on both the transverse momentum and the measured charged-particle multiplicity.  At low multiplicity, the ratio measured at LHCb is consistent with the value measured in $e^{+}e^{-}$ collisions, and increases by a factor of $\sim2$ with increasing multiplicity.  At relatively low transverse momentum, the ratio of $\Lb$ to $\Bd$ cross-sections is higher than what is measured in $e^{+}e^{-}$ collisions, but converges with the $e^{+}e^{-}$ ratio as the momentum increases.  These results imply that the evolution of heavy $b$ quarks into final-state hadrons is influenced by the density of the hadronic environment produced in the collision.  Comparisons with several models and implications for the mechanisms enforcing quark confinement are discussed.
  
\end{abstract}

\vspace*{2.0cm}

\begin{center}
Published in Phys.~Rev.~Lett. 132 (2024) 081901
\end{center}

\vspace{\fill}

{\footnotesize 
\centerline{\copyright~\papercopyright. \href{\paperlicenceurl}{\paperlicence}.}}
\vspace*{2mm}

\end{titlepage}


\newpage
\setcounter{page}{2}
\mbox{~}
%
%
%
%


\renewcommand{\thefootnote}{\arabic{footnote}}
\setcounter{footnote}{0}

\cleardoublepage


\pagestyle{plain} 
\setcounter{page}{1}
\pagenumbering{arabic}


One of the defining features of quantum chromodynamics (QCD) is color confinement, which prohibits isolated quarks and gluons from being observed.  Instead, the partons, which carry color charge, are found only as constituents of color-neutral hadrons~\cite{Wilson:1974sk}. Quarks that are produced at colliders such as the Large Hadron Collider (LHC), evolve into observable hadrons through a process known as hadronization.  Factorization theorems in QCD typically assume that hadronization is a universal process, independent of the colliding beam species~\cite{Collins:1989gx}. 

One mechanism of hadronization is parton-to-hadron fragmentation, where the potential between outgoing partons grows until it is energetically favorable to produce other partons from the vacuum that neutralize the initiating parton's color charge.  This process is often modeled by cluster formation~\cite{Andersson:1983ia} or color string breaking~\cite{Webber:1983if}. The resulting collections of particles are known as jets~\cite{Field:1977fa}, and have been studied extensively in $e^{+}e^{-}$, $pp$, $p\bar{p}$, and heavy-ion collisions~\cite{Barber:1979yr, JADE:1986kta, Mueller:1990gy, CDF:1996yow, D0:2008nou, Bhatti:2010bf, Kogler:2018hem, Connors:2017ptx}.  Fragmentation functions, which describe the evolution of quarks and gluons resulting from hard scattering into hadrons, are constrained by data from $e^{+}e^{-}$ collisions~\cite{Kniehl:2000hk,Metz:2016swz,deFlorian:2007aj}.

An alternative hadronization mechanism can arise when quarks that are produced near each other combine to form color-neutral hadrons through a process called coalescence. Models of coalescence generally require individual parton wave functions to overlap in position and velocity space, and have successfully described a range of collider and fixed-target data~\cite{Das:1977cp,Vogt:1994zf,Braaten:2002yt,Rapp:2003wn,He:2019vgs}.  In these models, the density of partons produced in the underlying event has a significant effect on the hadronization process, and is expected to be especially important in heavy-ion collisions~\cite{Fries:2003kq,Rapp:2003wn}.  Enhanced baryon production can be a signature of coalescence, and models generally require coalescence in order to reproduce the baryon/meson production cross-section ratios measured in heavy-ion collisions~\cite{Hwa:2002tu,Fries:2003vb,Greco:2003xt}.  Multiple parton interactions in $pp$ collisions can also result in many overlapping color strings and high final-state particle multiplicities. This alternative hadronization mechanism does not require fragmentation, and since charged-particle multiplicity in the underlying event depends on the colliding beam species, the emergence of coalescence in hadron collisions breaks the universality of the hadronization process.

The family of hadrons containing bottom quarks is particularly sensitive to hadronization. Because of the large mass of $b$ quarks, light quark to $b$-hadron fragmentation is suppressed, and the long wavelengths of heavy $b$ quarks at low transverse momentum $\pt$ can potentially overlap with other particles produced in the event.  Additionally, there is no $b$-quark content in the valence quarks of beam particles, and $b$-quark production is dominated by hard parton-parton interactions in the early stages of the collision, before hadronization takes place~\cite{Norrbin:2000zc}.  Total $b\bar{b}$ cross-sections are consequently well described by perturbative QCD calculations~\cite{LHCb-PAPER-2016-031,LHCb-PAPER-2017-037,ALICE:2021mgk}.  Previous measurements have shown that the fraction of $b$ quarks that hadronize into baryons has a significant $\pt$ dependence~\cite{LHCb-PAPER-2011-018, LHCb-PAPER-2018-050}, and that there is an asymmetry between $\Lb$ and $\Lbbar$ production at the LHC~\cite{LHCb-PAPER-2021-016}. There is also evidence for an increase in the rate of $\Bs$ meson production relative to $\Bd$ production with increasing multiplicity at low $\pt$~\cite{LHCb-PAPER-2022-001}. Additionally, the ALICE Collaboration has shown that the fragmentation fractions for charm quarks vary between $e^{+}e^{-}$ and $pp$ collisions, with $pp$ data favoring higher $\Lc$ baryon production~\cite{ALICE:2021dhb}, and that the baryon production is affected by the multiplicity of the underlying event~\cite{ALICE:2021npz}.  None of these results are expected in a scenario where heavy quarks hadronize only via fragmentation in the vacuum.  If coalescence is affecting $b$-quark hadronization, enhanced production of low-$\pt$ $\Lb$ baryons would be expected in collisions with relatively high multiplicity. In contrast, $b$ quarks produced in low multiplicity collisions and/or at high $\pt$ would have little interaction with other produced particles and fragment in the vacuum, as in $e^{+}e^{-}$ collisions.

This Letter presents new measurements of the ratio of $\Lb$ baryon to $\Bd$ meson production cross-sections, as a function of $\pt$ and multiplicity, using a sample of $pp$ collisions at $\sqrt{s} = 13\tev$ corresponding to an integrated luminosity of about 5.4 fb$^{-1}$. Candidates for $\Lb$ baryons and $\Bd$ mesons are reconstructed through their decays to the $\jpsi p K^{-}$ and $\jpsi \pi^{-} K^{+}$ final states, respectively (charge conjugation is assumed throughout this Letter), where the $\jpsi$ meson subsequently decays to the $\mup\mu^{-}$ final state.

The \lhcb detector~\cite{LHCb-DP-2008-001,LHCb-DP-2014-002} is a single-arm forward
spectrometer covering the \mbox{pseudorapidity} range $2<\eta <5$,
designed for the study of particles containing \bquark or \cquark quarks. The detector includes a high-precision tracking system
consisting of a silicon-strip vertex detector surrounding the $pp$
interaction region~\cite{LHCb-DP-2014-001} (known as the VELO detector), a large-area silicon-strip detector located
upstream of a dipole magnet with a bending power of about
$4{\mathrm{\,Tm}}$, and three stations of silicon-strip detectors and straw
drift tubes~\cite{LHCb-DP-2017-001} placed downstream of the magnet.
The tracking system provides a measurement of the momentum of charged particles with
a relative uncertainty that varies from 0.5\% at low momentum to 1.0\% at 200\gevc.  Different types of charged hadrons are distinguished using information
from two ring-imaging Cherenkov detectors~\cite{LHCb-DP-2012-003}. Muons are identified by a system composed of alternating layers of iron and multiwire proportional chambers~\cite{LHCb-DP-2012-002}. 

Simulation is required to model the effects of the detector acceptance and the imposed selection requirements.  In the simulation, $pp$ collisions are generated using \pythia~\cite{Sjostrand:2007gs}.  Decays of unstable particles are described by \evtgen~\cite{Lange:2001uf}, in which final-state radiation is generated using \photos~\cite{davidson2015photos}. The interaction of the generated particles with the detector, and its response, are implemented using the \geant toolkit~\cite{Allison:2006ve} as described in Ref.~\cite{LHCb-PROC-2011-006}. Weights are applied to the simulation to ensure the invariant mass distributions of product pairs from the $\Lb$ and $\Bd$ decays and the parent $\pt$ match background-subtracted distributions that are extracted from the data using the \sPlot method~\cite{splot}. 

Events considered here are identified by a series of hardware and software triggers designed to select muons from the decay $\jpsi\rightarrow\mup\mu^{-}$, and have exactly one reconstructed primary vertex. Each individual muon candidate is required to have $\pt > 500 \mevc$ and to penetrate hadron absorbers in the LHCb muon system.  The reconstructed $\mup\mu^{-}$ pair must have a mass within $39 \mevcc$ of the known $\jpsi$ mass (which corresponds to 3 times the measured mass resolution of the $\jpsi$ peak)~\cite{PDG2020}, and must form an origin vertex that is displaced from the primary vertex. Charged pion, kaon and proton candidates are positively identified by the response of the LHCb ring-imaging Cherenkov detectors, and are required to have a momentum above $3 \gev/c$ and $\pt > 750 \mev/c$. Combinations of tracks that meet these requirements are refit with kinematic constraints that fix the $\mup\mu^{-}$ invariant mass to the known $\jpsi$ mass, and require all four tracks to share a common vertex. 

The multiplicity metrics used are either the total number of charged tracks reconstructed in the VELO detector, $\Nvelo$, or the subset of backwards VELO tracks that point away from the LHCb spectrometer, $\Nback$, which covers the pseudorapidity interval of approximately $-3.5 \ < \eta < -1.5$. The backwards VELO tracks provide a multiplicity estimate that is measured in a different rapidity region than the signal. Events containing signal candidates are characterized in terms of their relation to events that are selected by triggering on the LHC beam clock alone, no bias (NB) events, which provide a selection of $pp$ collisions that do not have biases resulting from trigger decisions based on measured particles.  The results are quoted in terms of normalized multiplicity, defined as the number of tracks at the center of a given multiplicity bin divided by the mean number of tracks in NB events, which are $\langle \Nvelo \rangle$$_{\mathrm{NB}}$ = 37.7 and $\langle \Nback \rangle$$_{\mathrm{NB}}$ = 11.1, with negligible statistical uncertainties of less than $0.1\%$~\cite{LHCb-PAPER-2022-001}.

\begin{figure}[bt]
  \centering
  \includegraphics[width=1.0\linewidth]{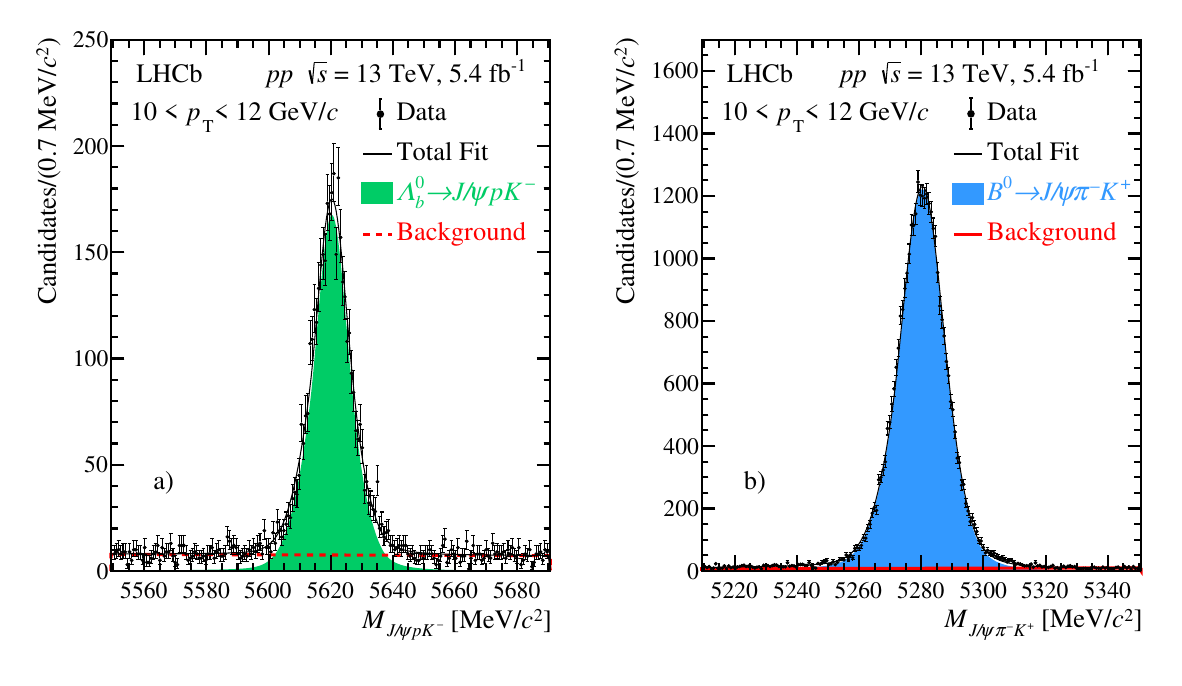}
  \caption{Invariant mass spectra in the transverse momentum interval $10<\pt<12 \gevc$ for the (a) $\jpsi p K^{-}$ and (b) $\jpsi \pi^{-} K^{+}$ final states, integrated over multiplicity. Fit projections including signal and background components are shown.} 
  \label{fig:Fig1}
\end{figure}

Extended maximum-likelihood fits are performed on the binned $\jpsi p K^{-}$ and $\jpsi \pi^{-} K^{+}$ invariant mass spectra to determine the $\Lb$ and $\Bd$ yields, respectively, in intervals of $\pt$ and multiplicity.  The $\Lb$ and $\Bd$ signals are each described by a sum of two Crystal Ball functions~\cite{Skwarnicki:1986xj}, whose tail parameters and relative contributions are fixed to values determined by fitting simulated decays. The background contribution is represented by an exponential function, which provides a good description of the mass sidebands. Examples of the $\jpsi p K^{-}$ and $\jpsi \pi^{-} K^{+}$ invariant mass distributions and fits are shown in Fig.~\ref{fig:Fig1}.  Variations of the fit functions that converge change the extracted $\Lb$ and $\Bd$ yields by about $1\%$, which is taken as a systematic uncertainty.

The ratio of production cross-sections $\sigma_{\Lb}/\sigma_{\Bd}$ is found by calculating 

\begin{equation}
\frac{\sigma_{\Lb}}{\sigma_{\Bd}} = \frac{N_{\Lb}}{N_{\Bd}} \times \frac{\BR_{\Bd}}{\BR_{\Lb}} \times \frac{\epsilon^{\rm acc}_{\Bd}}{\epsilon^{\rm acc}_{\Lb}} \times \frac{\epsilon^{\rm trig}_{\Bd}}{\epsilon^{\rm trig}_{\Lb}} \times \frac{\epsilon^{\rm rec}_{\Bd}}{\epsilon^{\rm rec}_{\Lb}} \times \frac{\epsilon^{\rm PID}_{\Bd}}{\epsilon^{\rm PID}_{\Lb}},
\end{equation} 
where $N_{\Lb}$ and $N_{\Bd}$ are the $\Lb$ and $\Bd$ yields extracted by fitting the peaks in the invariant mass spectra,  $\BR_{\Lb}$ and $\BR_{\Bd}$ are the branching fractions of $\Lb$ and $\Bd$ to the relevant final states~\cite{PDG2020}, $\epsilon^{\rm acc}$ is the acceptance of the LHCb spectrometer for the given decay, $\epsilon^{\rm trig}$ is the efficiency for triggering on the given decay, $\epsilon^{\rm rec}$ is the efficiency for selecting and reconstructing the parent particle, and $\epsilon^{\rm PID}$ is the total particle identification efficiency for the final-state charged tracks from the decays.

The ratio leads to the partial cancellation of many of the correction factors and their systematic uncertainties due to the similarities of the decays. The ratio of acceptance corrections $\epsilon^{\rm acc}_{\Bd} / \epsilon^{\rm acc}_{\Lb}$ is estimated via simulation to be $\sim0.95$ at low $\pt$ and tends to unity with increasing $\pt$. A systematic uncertainty of $1\%$ is evaluated by varying the weights applied to the simulated events within their uncertainties.  The ratio of trigger efficiencies $\epsilon^{\rm trig}_{\Bd} / \epsilon^{\rm trig}_{\Lb}$ is measured as described in Ref.~\cite{LHCb-PUB-2014-039}, and found to be consistent with unity. Uncertainties on this factor vary from 1$\%$ at low $\pt$ to $3\%$ at high $\pt$, due to the finite size of the data samples used to measure the trigger efficiencies. The ratio of reconstruction efficiencies $\epsilon^{\rm rec}_{\Bd} / \epsilon^{\rm rec}_{\Lb}$ varies from $\sim0.8$ at the lowest $\pt$ values and increases to be consistent with unity where $\pt$ is much greater than the $\Lb$ and $\Bd$ masses.  The deviation from unity is due to the differences in the  reconstruction efficiencies of the intermediate states in the products of the $\Lb$ and $\Bd$ decays~\cite{LHCb-PAPER-2015-029,LHCb-PAPER-2018-043,LHCb-PAPER-2019-014}. A systematic uncertainty that varies with $\pt$ from $2\%$ to $4\%$ is assigned to account for uncertainties on the weights applied to the simulation, and an additional uncertainty that is negligible at low $\pt$ and increases to a maximum of $\sim2\%$ at high $\pt$ is assigned to account for the finite size of the simulation samples used to calculate this correction.  The ratio of particle identification efficiencies $\epsilon^{\rm PID}_{\Bd} / \epsilon^{\rm PID}_{\Lb}$ is found using calibrated samples of identified particles taken from data. This term deviates from unity by a maximum of 7$\%$ at low $\pt$, due to the difference in identification efficiencies for pions from $\Bd$ decays and protons from $\Lb$ decays.  A systematic uncertainty of $2\%$ is assigned due to the finite size of the calibration samples~\cite{LHCb-PUB-2016-021}. The product of the efficiency ratios ranges from $\sim0.8$ at low $\pt$ to $\sim0.9$ at high $\pt$.  The systematic uncertainty on these corrections is calculated by adding all contributions in quadrature, and ranges between $4\%$ and $6\%$. Systematic uncertainties dominate for $\pt<18\gevc$, with statistical uncertainties becoming dominant at higher $\pt$.
  


The resulting $\sigma_{\Lb}/\sigma_{\Bd}$ ratio is shown in Fig.~\ref{fig:Fig2} as a function of $\pt$.  Numerical values are given in the Supplemental Material~\cite{ref:SuppMat}.  The blue points show the data analyzed in this Letter, where the error bars are the quadrature sum of the statistical and systematic uncertainties, and the error boxes indicate the point-to-point fully correlated (global) uncertainty  due to uncertainties on the $\Lb$ and $\Bd$ branching fractions. The gray points show a previous measurement of the ratio of the $\Lb$ cross-section to the sum of the $\Bd$ and $\Bu$ cross-sections measured using semileptonic decays over the pseudorapidity interval $2<\eta<5$~\cite{LHCb-PAPER-2018-050}. For direct comparison here, that ratio is multiplied by a factor of 2 under the assumption of equal production of $\Bd$ and $\Bu$ mesons, which is supported by data~\cite{LHCb-PAPER-2013-004}.  In the $\pt$ range where the measurements overlap, there is good agreement within uncertainties. Data from $p$Pb collisions~\cite{LHCb-PAPER-2018-048} covering two different rapidity intervals are also shown for comparison (purple points), and generally agree with the $pp$ data within uncertainties.

The solid green and dashed dark green curves in Fig.~\ref{fig:Fig2} are from a recent model of statistical hadronization of $b$ quarks~\cite{He:2022tod} that assumes relative chemical equilibrium between different $b$-hadron yields, and considers two sets of $b$ hadrons as input. The green solid curve uses the measured spectrum of baryons collected by the Particle Data Group (PDG) in Ref.~\cite{PDG2020}, while the dark dashed curve uses an expanded set of $b$ baryons that are expected by the relativistic quark model (RQM)~\cite{Ebert:2011kk}. Feeddown contributions from these baryons would contribute to the $\Lb$ yields. The central values of the data are most consistent the RQM calculation at intermediate $\pt$, while the PDG calculation is favored at the lowest and highest $\pt$. 
    
Results from two event generators are also shown in Fig.~\ref{fig:Fig2}.  The black dashed line is a calculation from PYTHIA8, using the default settings.  This model shows little variation with $\pt$, but is consistent with the data for $\pt>20 \gevc$.  Results from the EPOS4HQ event generator \cite{Zhao:2023ucp} are given for two configurations. The dashed orange curve shows the ratio where all $b$ quarks are required to hadronize via fragmentation, which is largely consistent with the result from PYTHIA8, and does not match the data at low $\pt$.  The other EPOS4HQ calculation, shown in the red dashed line (``EPOS4HQ+coal"), includes quark coalescence as an additional hadronization mechanism for $b$ quarks.  This model reproduces the shape of the data, but slightly overpredicts the magnitude of the $\sigma_{\Lb}/ \sigma_{\Bd}$ ratio at intermediate $\pt$.  This could indicate that quark coalescence plays a role in baryon formation at relatively low $\pt$, while high-$\pt$ $b$ hadrons are formed primarily through fragmentation in vacuum. 


 \begin{figure}[!ht]
  \centering
  \includegraphics[width=1.0\linewidth]{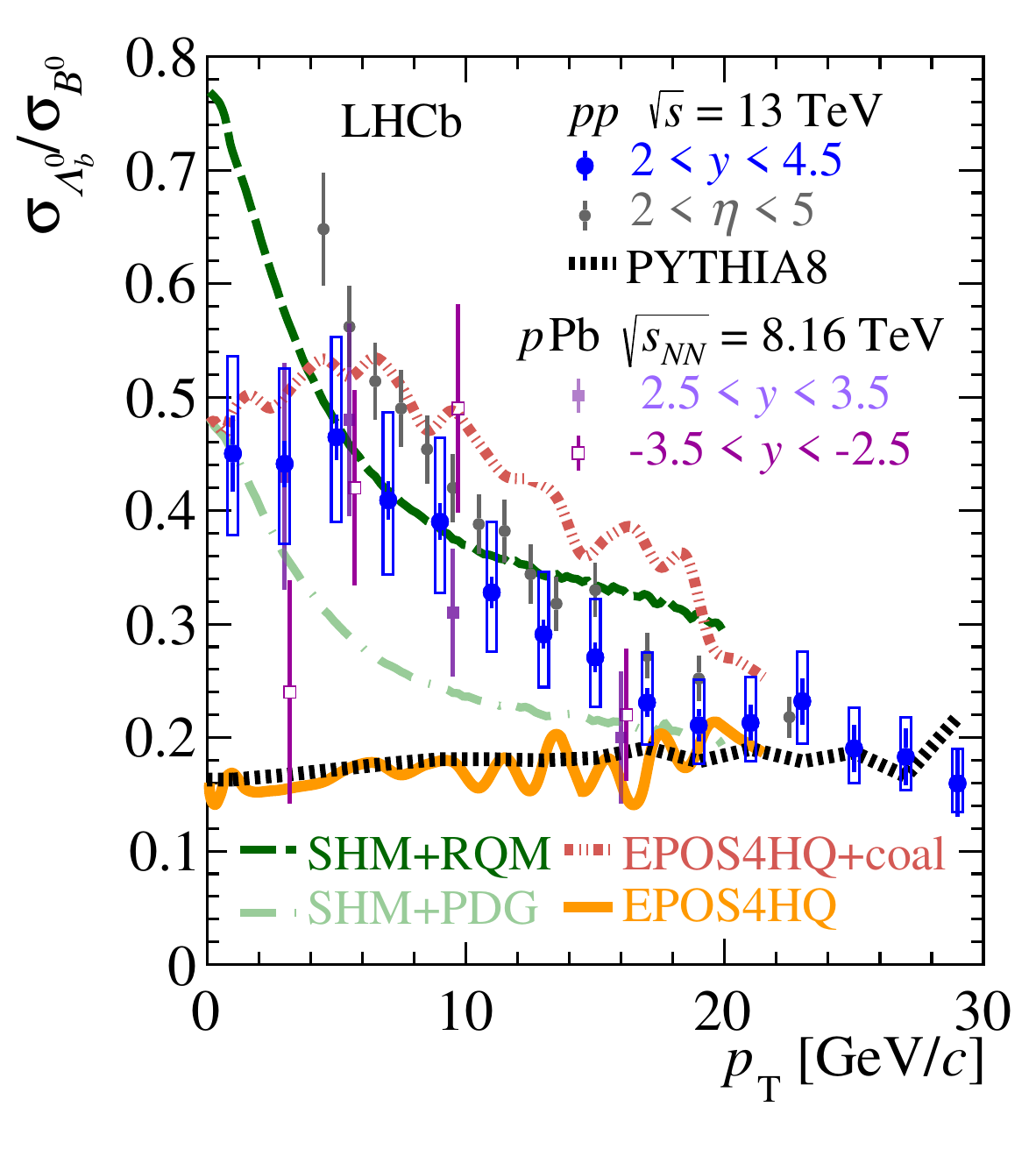}
  \caption{Ratio of $\Lb$ to $\Bd$ production cross-sections as a function of $\pt$ (blue circles).  The error bars represent the quadrature sum of the statistical and systematic uncertainties and the boxes depict the global uncertainties due to branching fractions. The data are compared to previous $pp$ measurements using semileptonic decays (gray circles)~\cite{LHCb-PAPER-2018-050}, and data from $p$Pb collisions (squares)~\cite{LHCb-PAPER-2018-048}.  Calculations from a statistical hadronization model (SHM)~\cite{He:2022tod}, and the PYTHIA8 \cite{Sjostrand:2007gs} and EPOS4HQ \cite{Zhao:2023ucp} event generators are also shown.} 
  \label{fig:Fig2}
\end{figure}

The $\pt$-integrated $\sigma_{\Lb}/ \sigma_{\Bd}$ ratio is shown as a function of multiplicity in Fig.~\ref{fig:Fig3}. Numerical values are given in the Supplemental Material~\cite{ref:SuppMat}.  The error bars represent the quadrature sum of the statistical and systematic uncertainties, while the global uncertainty of $\ ^{+19\%}_{-16\%} $ due to the branching fractions is indicated in text. In the lowest multiplicity bins, this $pp$ data approaches the baryon fraction measured in $e^{+}e^{-}$ collisions at the Large Electron-Position collider (LEP)~\cite{HFLAV18}. This is expected since $b$ quarks in low-multiplicity events do not overlap with other particles and fragment in the vacuum, as in $e^{+}e^{-}$ collisions.  There is a distinct rise of the baryon fraction with multiplicity, which plateaus at $\sim0.5$ for collisions that produce more than twice the average number of VELO tracks, though the data at the highest multiplicities have increasing uncertainties.  This could indicate that coalescence emerges as an additional production mechanism for baryons at high multiplicity, where multiple quark wave functions overlap.

 \begin{figure}[!ht]
  \centering
  \includegraphics[width=1.0\linewidth]{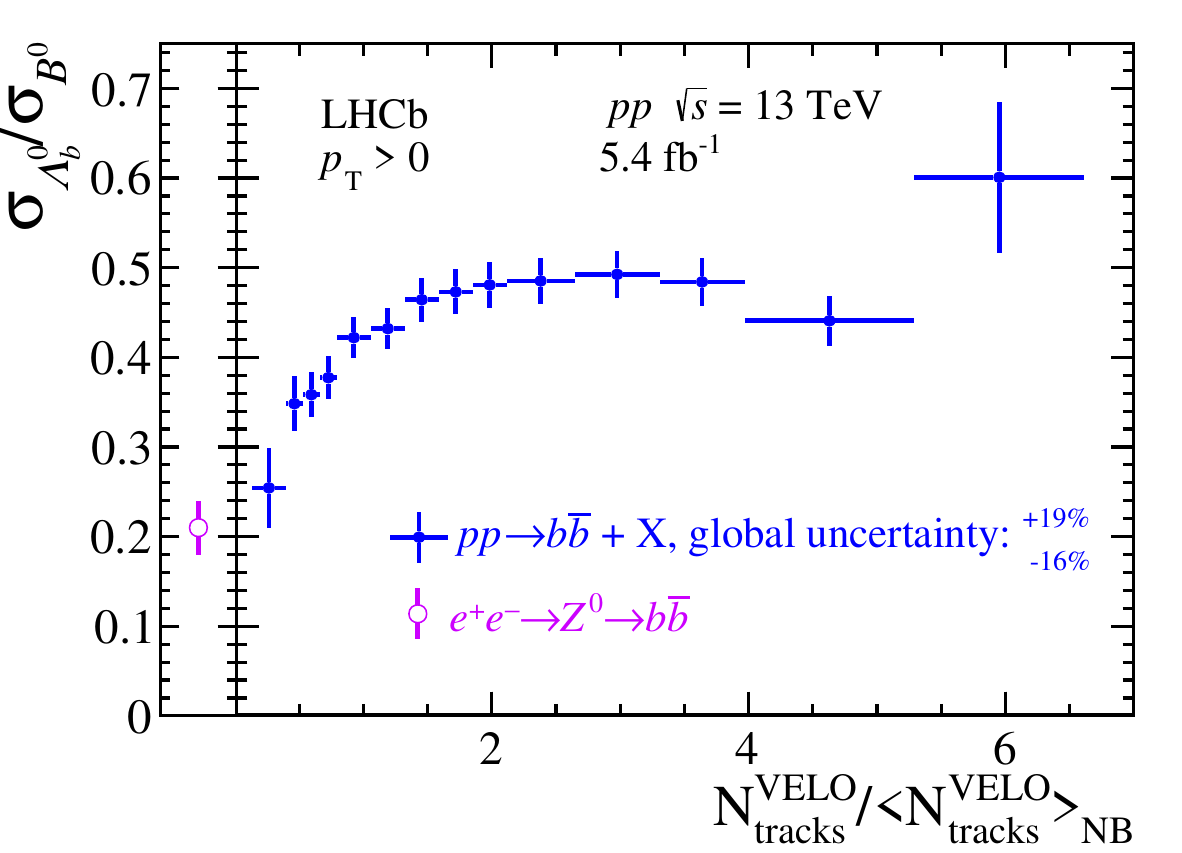}
  \caption{Ratio of $\Lb$ to $\Bd$ cross-sections as a function of the total track multiplicity measured in the VELO detector (blue).  The purple point indicates the value measured in $e^{+}e^{-}$ collisions at LEP~\cite{HFLAV18}. } 
  \label{fig:Fig3}
\end{figure}

The $\sigma_{\Lb}/ \sigma_{\Bd}$ ratio is shown in Fig.~\ref{fig:Fig4} as a function of $\pt$ in different multiplicity bins. Numerical values are given in the Supplemental Material~\cite{ref:SuppMat}.  The left panel shows the data in the different bins of total track multiplicity, where the multiplicity intervals correspond to less than $\langle \Nvelo \rangle$$_{\mathrm{NB}}$, between 1 and 2 times $\langle \Nvelo \rangle$$_{\mathrm{NB}}$, and greater than twice $\langle \Nvelo \rangle$$_{\mathrm{NB}}$. The right panel shows similar data, but uses $\langle \Nback \rangle$$_{\mathrm{NB}}$ to define multiplicity. The error bars represent the quadrature sum of the statistical and systematic uncertainties, while the global uncertainty on all data points due to the uncertainties on the branching fractions is indicated in text. The purple point represents the LEP measurement of the ratio of the $b$-baryon fraction to $b$-meson fraction~\cite{HFLAV18}.

A significant multiplicity dependence can be observed using both multiplicity metrics, with higher multiplicity collisions displaying a larger $\Lb$ fraction.  The dependence is more pronounced when using the total number of VELO tracks (dominated by forward tracks), where there is clear ordering at intermediate $\pt$, however the backwards VELO tracks metric also shows a clear difference between the lowest and higher multiplicity bins.  At low $\pt$, the ratio $\sigma_{\Lb}/ \sigma_{\Bd}$ is significantly higher than the value measured in $e^{+}e^{-}$ collisions, but as $\pt$ increases the multiplicity dependence weakens and all data converges to the value from LEP.  This is expected in quark coalescence scenarios, where low-$\pt$ $\Lb$ baryons can be produced when $b$ quarks combine with two light quarks. However, at high-$\pt$ $\Lb$ baryons have limited overlap with the underlying event and are therefore produced by $b$-quark fragmentation in the vacuum.

 \begin{figure}[!ht]
  \centering
  \includegraphics[width=1.0\linewidth]{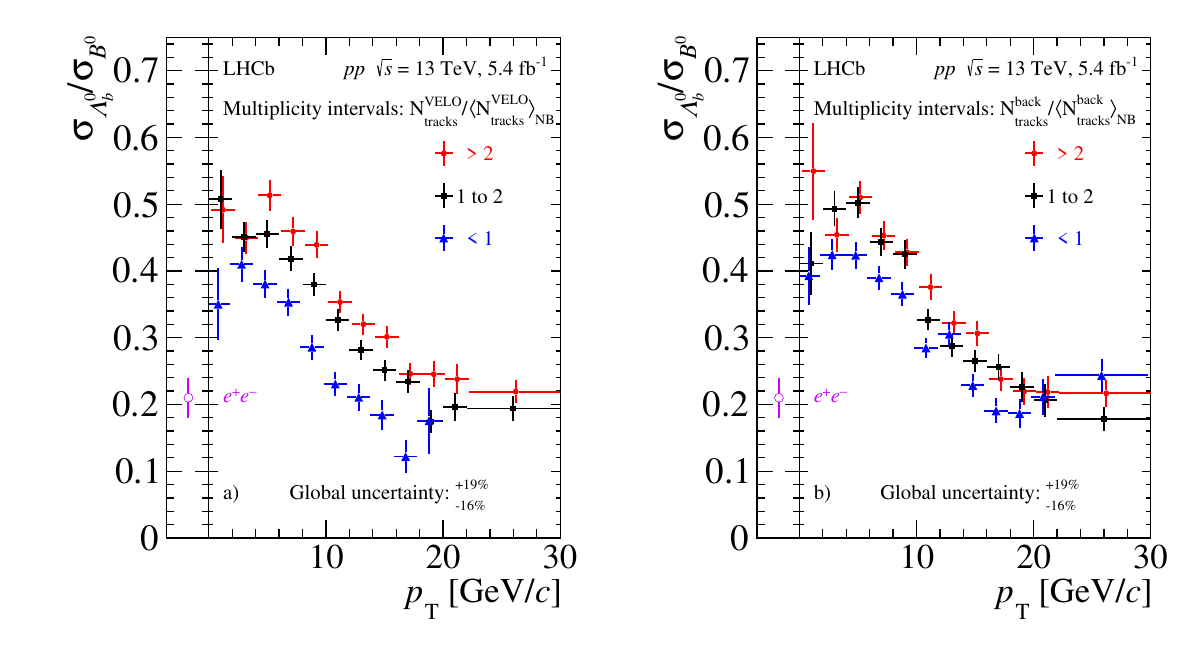}
  \caption{Ratio of $\Lb$ to $\Bd$ cross-sections as a function of $\pt$, in bins of a) the total multiplicity measured in the VELO detector and b) the backwards track multiplicity. The purple point shows the value measured in $e^{+}e^{-}\rightarrow\Z^{0}\rightarrow b\bar{b}$ reactions at LEP~\cite{HFLAV18}. } 
  \label{fig:Fig4}
\end{figure}

In summary, these new measurements from LHCb show clear variations of the $\sigma_{\Lb}/ \sigma_{\Bd}$ ratio with multiplicity and $\pt$ in $pp$ collisions at $\sqrt{s} = 13 \tev$.  Production of the $\Lb$ baryon is enhanced relative to the $\Bd$ meson at high charged-particle multiplicity and relatively low $\pt$. The baryon fraction measured at LEP can be reached in $pp$ collisions with very low multiplicity, and with $b$ hadrons at high $\pt$, regardless of multiplicity.  These observations show that $b$-quark hadronization is not universal between $e^{+}e^{-}$ and $pp$ collisions, and may indicate that quark coalescence plays an important role in forming hadrons at the LHC. These results are qualitatively consistent with the emergence of coalescence as an additional baryon production mechanism in high-density hadronic environments.


\section*{Acknowledgements}
%
%
\noindent We express our gratitude to our colleagues in the CERN
accelerator departments for the excellent performance of the LHC. We
thank the technical and administrative staff at the LHCb
institutes.
We acknowledge support from CERN and from the national agencies:
CAPES, CNPq, FAPERJ and FINEP (Brazil); 
MOST and NSFC (China); 
CNRS/IN2P3 (France); 
BMBF, DFG and MPG (Germany); 
INFN (Italy); 
NWO (Netherlands); 
MNiSW and NCN (Poland); 
MCID/IFA (Romania); 
MICINN (Spain); 
SNSF and SER (Switzerland); 
NASU (Ukraine); 
STFC (United Kingdom); 
DOE NP and NSF (USA).
We acknowledge the computing resources that are provided by CERN, IN2P3
(France), KIT and DESY (Germany), INFN (Italy), SURF (Netherlands),
PIC (Spain), GridPP (United Kingdom), 
CSCS (Switzerland), IFIN-HH (Romania), CBPF (Brazil),
Polish WLCG  (Poland) and NERSC (USA).
We are indebted to the communities behind the multiple open-source
software packages on which we depend.
Individual groups or members have received support from
ARC and ARDC (Australia);
Key Research Program of Frontier Sciences of CAS, CAS PIFI, CAS CCEPP, 
Fundamental Research Funds for the Central Universities, 
and Sci. \& Tech. Program of Guangzhou (China);
Minciencias (Colombia);
EPLANET, Marie Sk\l{}odowska-Curie Actions, ERC and NextGenerationEU (European Union);
A*MIDEX, ANR, IPhU and Labex P2IO, and R\'{e}gion Auvergne-Rh\^{o}ne-Alpes (France);
AvH Foundation (Germany);
ICSC (Italy); 
GVA, XuntaGal, GENCAT, Inditex, InTalent and Prog.~Atracci\'on Talento, CM (Spain);
SRC (Sweden);
the Leverhulme Trust, the Royal Society
 and UKRI (United Kingdom).




\addcontentsline{toc}{section}{References}
\bibliographystyle{LHCb}
\bibliography{main,standard,LHCb-PAPER,LHCb-CONF,LHCb-DP,LHCb-TDR}
 
\newpage
\centerline
{\large\bf LHCb collaboration}
\begin
{flushleft}
\small
R.~Aaij$^{35}$\lhcborcid{0000-0003-0533-1952},
A.S.W.~Abdelmotteleb$^{54}$\lhcborcid{0000-0001-7905-0542},
C.~Abellan~Beteta$^{48}$,
F.~Abudin{\'e}n$^{54}$\lhcborcid{0000-0002-6737-3528},
T.~Ackernley$^{58}$\lhcborcid{0000-0002-5951-3498},
B.~Adeva$^{44}$\lhcborcid{0000-0001-9756-3712},
M.~Adinolfi$^{52}$\lhcborcid{0000-0002-1326-1264},
P.~Adlarson$^{78}$\lhcborcid{0000-0001-6280-3851},
C.~Agapopoulou$^{46}$\lhcborcid{0000-0002-2368-0147},
C.A.~Aidala$^{79}$\lhcborcid{0000-0001-9540-4988}
% ...

{\footnotesize \it

% ...
$^{35}$Nikhef National Institute for Subatomic Physics, Amsterdam, Netherlands\\
% ...
$^{44}$Instituto Galego de F{\'\i}sica de Altas Enerx{\'\i}as (IGFAE), Universidade de Santiago de Compostela, Santiago de Compostela, Spain\\
% ...
$^{46}$European Organization for Nuclear Research (CERN), Geneva, Switzerland\\
% ...
$^{48}$Physik-Institut, Universit{\"a}t Z{\"u}rich, Z{\"u}rich, Switzerland\\
% ...
$^{52}$H.H. Wills Physics Laboratory, University of Bristol, Bristol, United Kingdom\\
% ...
$^{54}$Department of Physics, University of Warwick, Coventry, United Kingdom\\
% ...
$^{57}$School of Physics and Astronomy, University of Glasgow, Glasgow, United Kingdom\\
$^{58}$Oliver Lodge Laboratory, University of Liverpool, Liverpool, United Kingdom\\
% ...
$^{66}$Syracuse University, Syracuse, NY, United States\\
% ...
$^{78}$Department of Physics and Astronomy, Uppsala University, Uppsala, Sweden, associated to $^{57}$\\
$^{79}$University of Michigan, Ann Arbor, MI, United States, associated to $^{66}$\\
% ...
}
\end{flushleft}

\end{document}